\documentclass[prd,showpacs,letterpaper,twocolumn]{revtex4}%
\usepackage{graphicx}
\usepackage{bm}
\usepackage{epsf}
\usepackage{rotating}
\usepackage{epsfig,graphics,rotate,color}
\usepackage{wrapfig}
\usepackage{amssymb}
\usepackage{amsmath}
\usepackage{amsfonts}
\usepackage{array,hhline,dcolumn}%
\providecommand{\U}[1]{\protect\rule{.1in}{.1in}}
\begin{document}
\title{Proposal for an Electron Antineutrino Disappearance Search Using High-Rate $^8$Li Production and Decay}

\author{A. Bungau$^1$, A. Adelmann$^2$, J.R. Alonso$^{3}$, W. Barletta$^{3}$, R. Barlow$^{1}$, L. Bartoszek$^{4}$, L. Calabretta$^{5}$, A. Calanna$^{3}$, D. Campo$^{3}$,
  J.M. Conrad$^{3}$, Z. Djurcic$^{6}$, Y. Kamyshkov$^{7}$, M.H. Shaevitz$^{8}$, I. Shimizu$^9$,
  T. Smidt$^3$, J. Spitz$^3$,  M. Wascko$^{10}$, L.A. Winslow$^{3}$, J.J. Yang$^{2,3}$}

\affiliation{$^1$ University of Huddersfield, Huddersfield, HD1 3DH, UK}
\affiliation{$^2$ Paul Scherrer Institut, Villigen, CH-5232, Switzerland}
\affiliation{$^3$ Massachusetts Institute of Technology, Cambridge, MA 02139, USA}
\affiliation{$^4$ Bartoszek Engineering, Aurora, IL 60506, USA}
\affiliation{$^5$ Istituto Nazionale di Fisica Nucleare, Laboratori Nazionali del Sud,
I-95123, Italy}
\affiliation{$^6$ Argonne National Laboratory, Argonne, IL 60439, USA}
\affiliation{$^7$ University of Tennessee, Knoxville, TN 37996, USA}
\affiliation{$^8$ Columbia University, New York, NY 10027, USA}
\affiliation{$^9$ Tohoku University, Sendai, 980-8578, Japan}
\affiliation{$^{10}$ Imperial College London, London, SW7 2AZ, United Kingdom}

\begin{abstract}
This paper introduces an experimental probe of the sterile neutrino with a novel, high-intensity source of electron antineutrinos from the production and subsequent decay of $^8$Li. When
paired with an existing $\sim$1~kton scintillator-based detector, this $\langle E_\nu\rangle$=6.4~MeV source opens a wide range of possible searches for beyond standard model physics via studies of the inverse beta decay interaction $\bar \nu_e + p \rightarrow  e^+ + n$. In particular, the experimental design described here has unprecedented sensitivity to $\bar \nu_e$ disappearance at $\Delta m^2\sim 1$~eV$^2$ and features the ability to distinguish between the existence of zero, one, and two sterile neutrinos. 
\end{abstract}

\maketitle
The beta decay-at-rest of $^8$Li produces an isotropic
electron antineutrino flux with an average energy of 6.4~MeV. An underground liquid scintillator based detector can be used to detect these antineutrinos via the inverse beta decay (IBD) process $\bar \nu_e + p
\rightarrow e^+ + n$. The antineutrino 
rate and energy, peaking at 9~MeV, can be
fully reconstructed by the detector. 
Precise energy and vertex reconstruction opens the possibility of searching for
antineutrino disappearance due to oscillations, which, in the simplest 
two-neutrino form, has the probability
\begin{equation}
P=1- \sin^2
2\theta \ \sin^2[1.27 \Delta m^2(L/E)]~,
\label{osc}
\end{equation}
where $\theta$ is the disappearance mixing angle; $\Delta m^2$ (eV$^2$) is the squared
mass splitting; $L$ is the distance (in meters) from the antineutrino source
to the detector; and $E$ (MeV) is the antineutrino energy. This probability is maximized
in the range of $\Delta m^2 \sim E/L$. An existing large scintillator-based antineutrino detector with a diameter of $\mathcal{O}$(10~m), when combined with an $^8$Li isotope
decay-at-rest source, is sensitive to oscillations at $\Delta m^2 \sim 1$
eV$^2$. This is an oscillation region of high interest due to 
anomalies that have been observed in the data from 
LSND~\cite{LSND}, MiniBooNE~\cite{MBnubar}, 
short-baseline reactor studies~\cite{Mention}, and gallium source
calibration runs~\cite{sagegallex}. These anomalies are often interpreted as being due
to sterile neutrinos~\cite{sorel, Karagiorgi, giunti, snac} and
have motivated the development of the IsoDAR (Isotope Decay-At-Rest) concept.

IsoDAR-style sources have been considered before~\cite{McDonald,
Russians, Russ2}. The design presented here, consisting of an ion source, cyclotron, and target, is the first with a sufficiently high antineutrino flux to address the existence of one or more sterile neutrinos. The $^9$Be target, used mainly as an intense source of beam-induced neutrons, is surrounded by a $^{7}$Li sleeve. When the target and sleeve combination is placed next to a kiloton-scale scintillator detector (e.g., KamLAND~\cite{KLdet}, SNO+~\cite{SNOplus}, or Borexino
\cite{Borexino}), the large antineutrino flux from $^8$Li beta decay can result in the collection of over 8$\times10^5$ IBD interactions in a five~year run. Such events allow a definitive search for antineutrino oscillations with the added ability to distinguish between models with one and two sterile neutrinos. A sample of more than 7200~$\bar \nu_e$-electron scatters is
also accumulated during this time and can be used as a sensitive electroweak probe.  

The charged particle beam, used for electron antineutrino production, originates with a 60~MeV/amu cyclotron accelerating 5~mA of H$_2^+$ ions.
The design of this compact cyclotron~\cite{injector} is ongoing and is envisaged as the injector for the accelerator system of the DAE$\delta$ALUS physics program~\cite{firstpaper, EOI}. 
The IsoDAR design calls for about a factor of six increase in intensity compared to compact cyclotrons used in the medical isotope industry. Current and future medical isotope machines accelerate protons to 60~MeV and beyond with average intensities of
$0.8-1.6$~mA~\cite{IBA,BEST}. 

In our design, a 5~mA H$_2^+$ beam is injected at 70~keV (35~keV/amu) via a spiral inflector. Existing ion sources are sufficient to supply the beam required~\cite{ion}. In the low-energy regime, space charge is crucial in modeling the beam dynamics correctly. The generalized perveance of a non-neutral beam, $K=qI/(2\pi\epsilon_0 m \beta^3\gamma^3$), parameterizes the strength of the space charge effect~\cite{perveance}. We conclude that the $K$ for this machine is of the same order as existing proton machine designs based on $\approx$2~mA and an injection energy of 30~keV~\cite{FRM2CCO04,adelmann}.The proposed high power is therefore feasible with regard to space charge issues.


The accelerator described is a continuous-wave source with a 90\% duty cycle to allow for machine maintenance. 
In consideration of target cooling and degradation with 600~kW of beam power, we require a uniform beam distributed across most of the 20~cm diameter target with a sharp cutoff at the edges. Third-order focussing elements in the extraction beam line are able to convert the Gaussian-like beam distribution into a nearly uniform one~\cite{Medas83} and hence create the necessary condition on the target.

\begin{figure}[h!]
\begin{center}
\begin{tabular}{c}
\hspace{-.5cm}
\includegraphics[scale=.25]{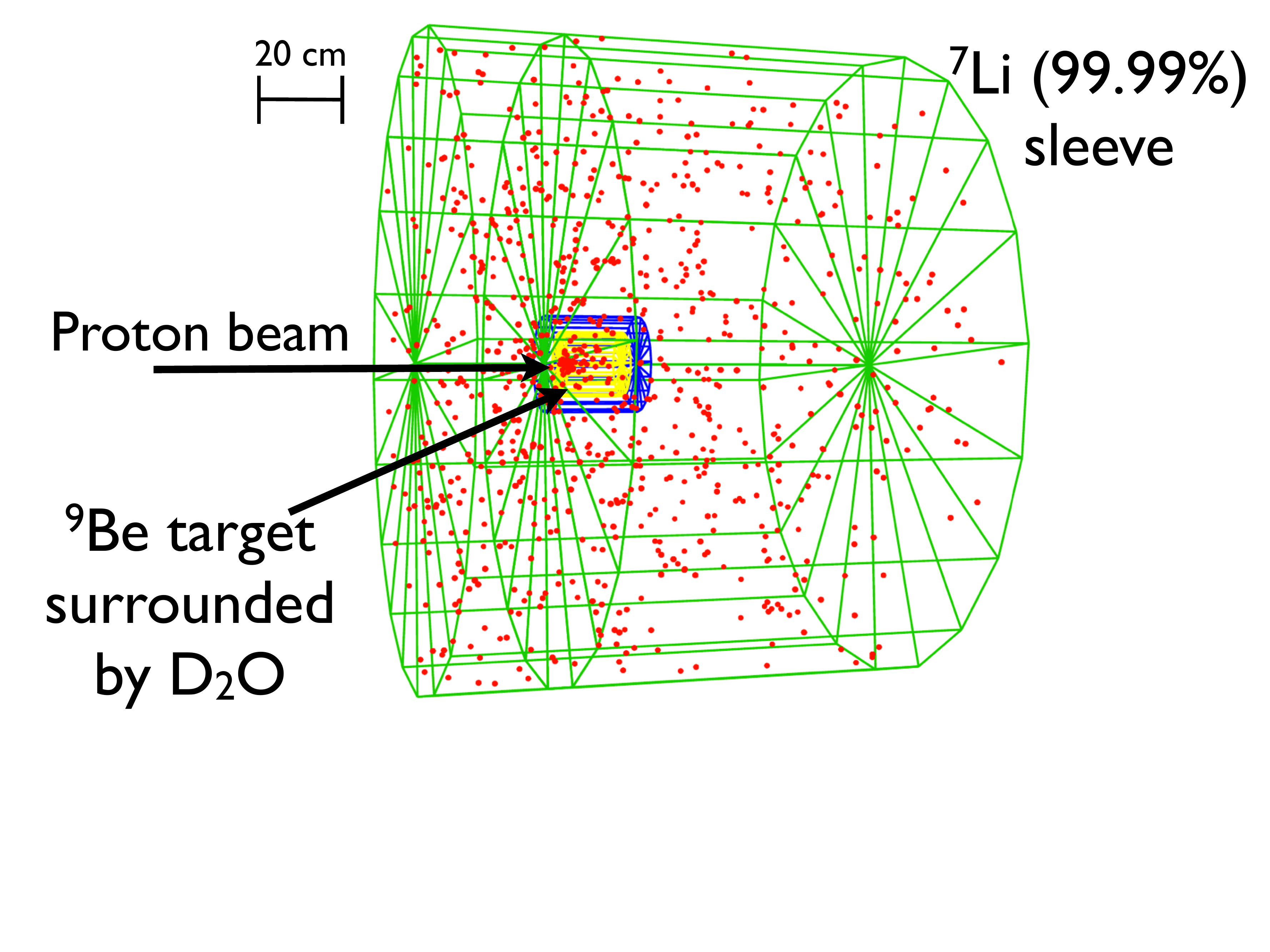}
\end{tabular}
\end{center}
\vspace{-1.9cm}
\caption{A schematic of the IsoDAR target and surrounding volumes. The dots represent $^8$Li ($\overline{\nu}_e$) creation points, obtained with $10^5$ 60~MeV protons on target simulated. The neutron reflector, shielding, and detector are not shown.}
\label{fig:schematic}
\end{figure}

The 60~MeV proton beam impinges on a cylindrical $^9$Be target that is
20~cm in diameter and 20~cm long. The primary purpose of this target 
is to provide a copious source of neutrons. Neutrons exiting the target are moderated and multiplied by a surrounding 5~cm thick region of 
D$_2$O, which also provides target cooling. Secondary neutrons enter a
150~cm long, 200~cm outer diameter cylindrical sleeve of 
solid lithium enveloping the target and D$_2$O layer. The target is embedded 40~cm into the upstream face of this volume; a window allows the beam to reach the target. The sleeve is composed of isotopically enriched lithium, 99.99\% $^7$Li compared to the natural abundance of 92.4\%. The isotopically pure material is widely used in the nuclear industry and is available from a number of sources. The isotope $^8$Li is formed by thermal neutron capture on $^7$Li and to a lesser extent by primary proton interactions in the $^9$Be target. For enhanced production, the sleeve is surrounded by a volume of graphite and steel acting as a neutron reflector and shield. The
volume extends 2.9~m in the direction of the detector. Isotope creation in the shielding is negligible. Figure~\ref{fig:schematic} displays the target and sleeve geometry, and Table~\ref{assumptions} summarizes the experimental parameters. We note that the geometry of the design is similar to that described in Ref.~\cite{Russians}.

\begin{table}[t]
\label{base_opti}
  \begin{center}
    {\footnotesize
      \begin{tabular}{|c|c|} \hline 
        Accelerator  & 60~MeV/amu of H$_2^+$  \\  \hline
        Current  & 10~mA of protons on target  \\  \hline
        Power  & 600~kW  \\  \hline
        Duty cycle  & 90\%  \\  \hline 
        Run period  & 5~years (4.5~years live time)  \\  \hline
        Target   & $^9$Be surrounded by $^7$Li (99.99\%)  \\  \hline
        $\overline{\nu}$ source  & $^8$Li $\beta$ decay ($\langle E_\nu\rangle$=6.4~MeV)  \\  \hline
        $\overline{\nu}_e$/1000 protons  &14.6 \\  \hline
        $\overline{\nu}_e$ flux  & 1.29$\times10^{23}$ $\overline{\nu}_e$ \\  \hline \hline
        Detector  & KamLAND   \\  \hline
        Fiducial mass  & 897 tons   \\  \hline
        Target face to detector center  & 16~m   \\  \hline
        Detection efficiency  & 92\%   \\  \hline
        Vertex resolution  & 12~cm/$\sqrt{E~\mathrm{(MeV)}}$ \\  \hline
        Energy resolution  & 6.4\%/$\sqrt{E~\mathrm{(MeV)}}$  \\  \hline
        Prompt energy threshold & 3 MeV  \\  \hline
        IBD event total & 8.2$\times 10^5$ \\  \hline
        $\overline{\nu}_e$-electron event total & 7200 \\ \hline 
      \end{tabular} 
      \caption{The relevant experimental parameters used in this study.}\label{assumptions}
}
\end{center}
\end{table}

We determine isotope production rates using a GEANT4 simulation~\cite{G4}. Due to its vast range of applications, GEANT4 provides an extensive set of data-based, parametrized, and theory-driven hadronic models, each one specializing in different types of interactions within a specified range of
energy. The QGSP-BIC-HP physics package was chosen for this particular application. The applicable physics model is the
pre-compound nuclear one which is invoked by the Binary Cascade simulation. Simulated hadronic processes include elastic scattering, inelastic
scattering, neutron capture, neutron fission, lepton-nuclear
interactions, capture-at-rest, and charge exchange. For neutron
energies below 20~MeV, the high-precision package uses the 
ENDF/B-VII data library~\cite{CSEWG}.

\begin{figure}[t]

{\includegraphics[width=3.4in]{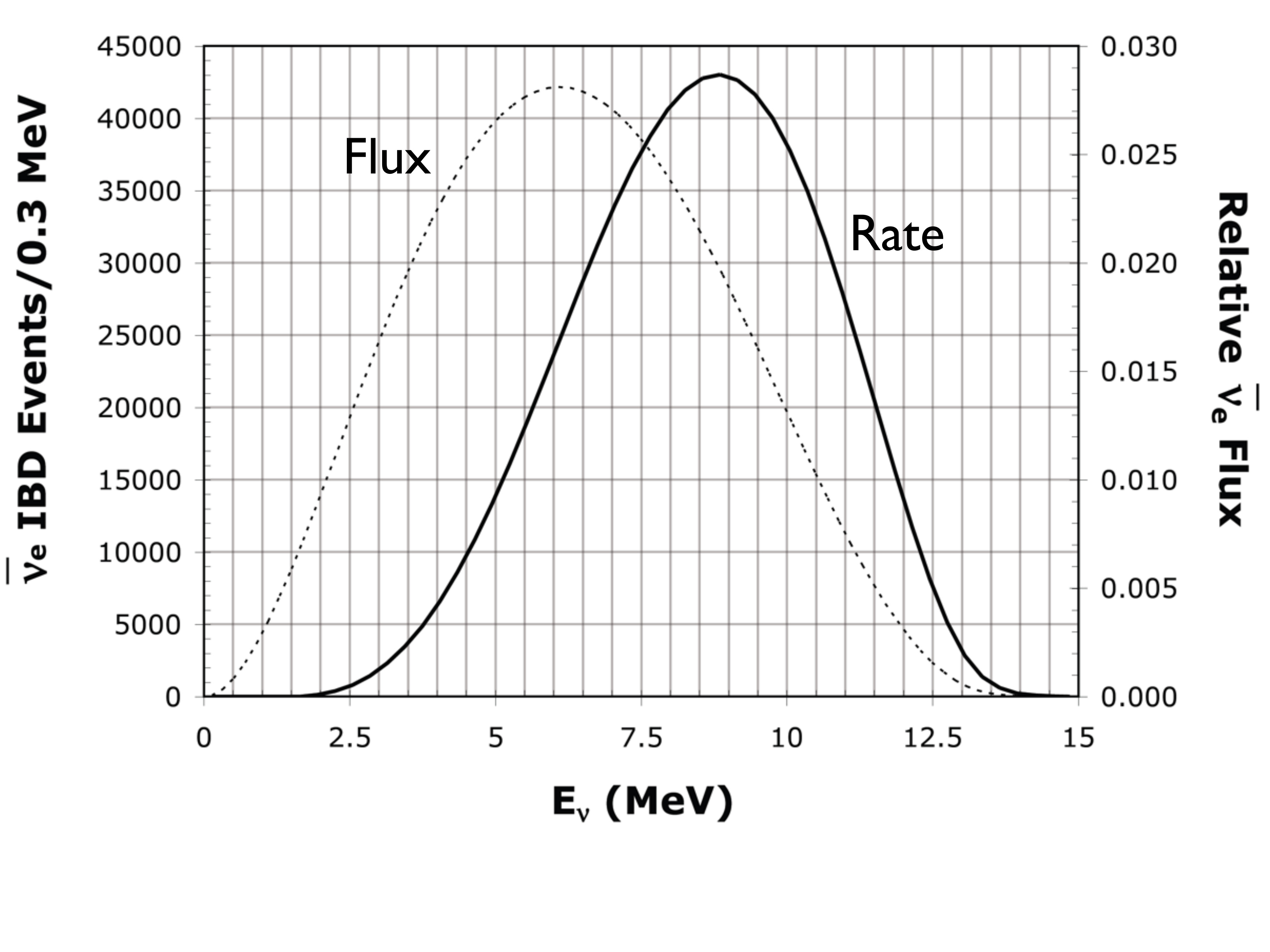}\hfill
}
\vspace{-1.3cm}
\caption{The expected antineutrino flux and detected event rate in the experimental configuration considered. The antineutrino flux mean energy from $^8$Li is 6.4~MeV. There are 8.2$\times 10^5$ reconstructed events expected from the 1.29$\times 10^{23}$~$\overline{\nu}_e$ created in the target and sleeve in five~years.
\label{spect}}
\end{figure}

Although all isotopes are considered in this analysis, the induced $^8$Li source in the sleeve dominates the antineutrino flux. The simulation yields 14.6~$^8$Li isotopes for every 1000 protons (60~MeV) on target. Approximately 10\% of all $^8$Li is produced inside the target; the rest is produced in the sleeve. Neutrinos and antineutrinos from other unstable isotopes are produced at a comparatively negligible rate. Over a five~year run period and with a 90\% duty cycle, 1.29$\times 10^{23}$ antineutrinos from $^8$Li are created. IsoDAR's nominal oscillation analysis is done in terms of ``shape-only" in $L/E$ and is therefore independent of the absolute flux normalization. However, a ``rate+shape" analysis using an absolute flux normalization uncertainty of 5\% is also considered in this study.

The IsoDAR antineutrino source is paired with an existing
underground scintillator-based detector for characterizing the antineutrino flux as a function of distance and energy. As can be seen in Eq.~\ref{osc}, a baseline of $L\sim10$~m is appropriate as a probe of the $\Delta m^2 \sim 1$~eV$^2$ region given the antineutrino spectrum shown in Fig.~\ref{spect}. We assume the face of the target sits 16~m from the center of a KamLAND-inspired 897~ton detector when calculating rates and oscillation sensitivity.

The $\bar \nu_e$ events are detected through the IBD interaction,  which is unique in several ways. IBD has a comparatively high cross section ($\sim 3\times 10^{-42}$~cm$^2$) which is known to $<1\%$ across the relevant energy range~\cite{VogelBeacom}. 
The distinct delayed coincidence signal of prompt light from the positron, followed by a 2.2~MeV gamma from the neutron capture on a proton, enables a low background rate expectation. The antineutrino energy can be fully reconstructed on an event-by-event basis using the visible energy of the prompt positron signal in the detector: $E_{\bar \nu_e}= E_{e^+} + 0.78$~MeV. A prompt energy threshold of 3~MeV is employed here. In a five year run, $8.2 \times 10^5$ reconstructed signal events are expected with 92\% detection efficiency~\cite{kamlandres} and in the absence of antineutrino oscillations.

The unique delayed coincidence signal makes reactor antineutrinos the only significant background in this analysis. The reactor antineutrino
event rate at KamLAND is on the order of 100 events per year
\cite{Kamland}, uniformly distributed across the detector. At SNO+, the reactor background would be a factor of about five lower~\cite{SNO+rate}. There is 9.4~m of passive and active shielding in between the end of the sleeve and the beginning of the fiducial volume, including an instrumented water veto detector. This shielding is adequate for attenuating/eliminating beam-related neutrons that can produce an IBD-like background, especially in consideration of the 3~MeV prompt energy threshold requirement. Furthermore, the sinusoidal-wave-like nature of an expected oscillation signal in $L/E$ cannot be mimicked by background.

\begin{figure*}[t]\begin{center}
{\includegraphics[width=7.05in]{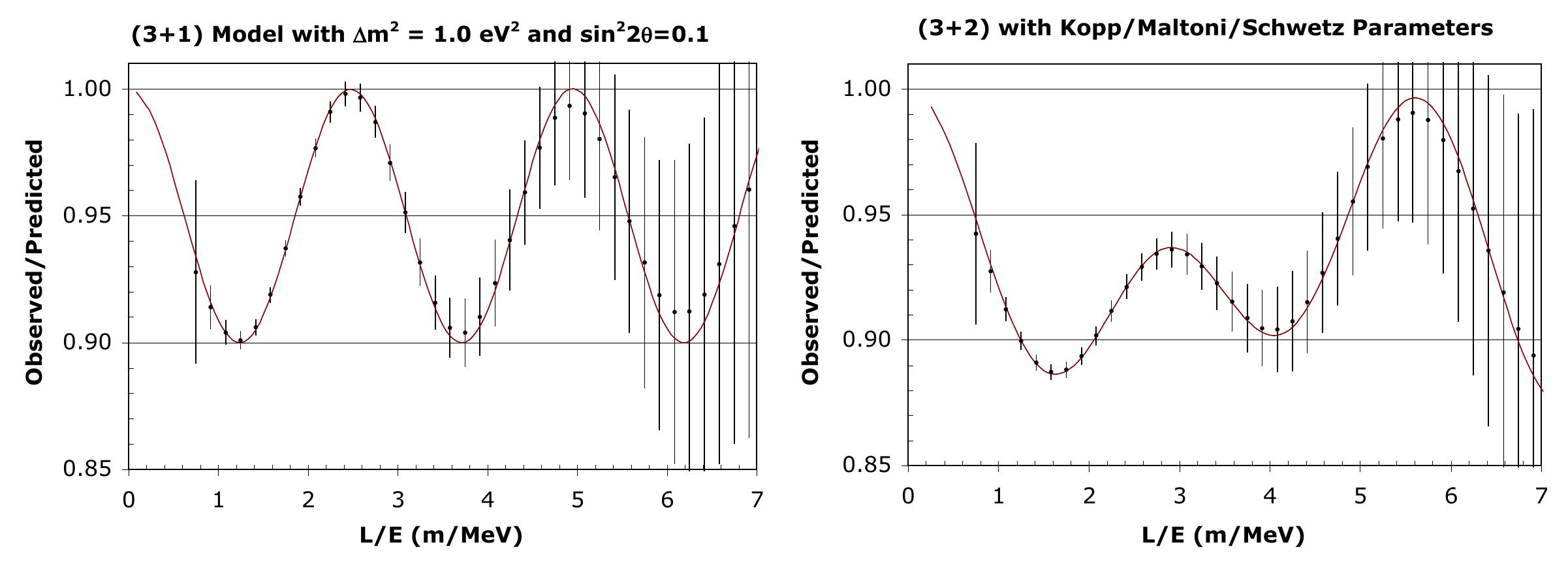}}
\end{center}
\vspace{-.5cm}
\caption{The $L/E$ dependence of two example oscillation
  signatures after five years of IsoDAR running. 
The solid curve is the oscillation probability with no smearing in the reconstructed position and energy. The 3+2 example (right) represents oscillations with the global best fit 3+2 parameters from Ref.~\cite{kopp}. 
\label{wiggles} }
\end{figure*}

To perform an oscillation analysis, the antineutrino travel distance ($L$) and energy ($E$) must be reconstructed simultaneously on an event-by-event basis.
Using KamLAND's detection capability as an example for the performance of 
a large scintillator detector,  the energy can be reconstructed with a
resolution of $6.4\%/\sqrt{E~\mathrm{(MeV)}}$~\cite{kamlandres}. With the antineutrino event vertex in the detector known to within 12~cm/$\sqrt{E~\mathrm{(MeV)}}$~\cite{kamlandres}, the $L$ resolution is dominated by
the spatial distribution of activated $^8$Li isotopes inside the target and sleeve. The antineutrino creation point is distributed in the beam coordinate $z$ according to an approximately uniform distribution, spanning the length of the 150~cm long sleeve. Although the spread in $z$ dominates the smearing of the antineutrino baseline $L$, the distribution in terms of the transverse coordinates is taken into account as well.

The IsoDAR oscillation analysis follows the method of Ref.~\cite{AgarwallaConradShaevitz}. This analysis exploits the $L/E$ dependence of oscillations, since $L$ and $E$ can be precisely reconstructed in the detector described. 
Eq.~\ref{osc} is a good approximation for 3+1 (three active plus one sterile) disappearance fits to data, but in the case of 3+2 (three active plus two sterile), Eq.~\ref{osc} is modified to accommodate $\Delta
m^2_{41}$, $\Delta m^2_{51}$, and $\Delta m^2_{45}$ oscillations. If $U$ is the mixing matrix, then the disappearance probabilities in the 3+1 and 3+2 scenarios are given by
\begin{eqnarray}
P_{3+1}&=&1-4|U_{e4}|^2(1-|U_{e4}|^2) \sin^2 (\Delta m^2_{41} L/E) \\
P_{3+2}&=&1-4[(1-|U_{e4}|^2-|U_{e5}|^2)\times  \nonumber \\
& & (|U_{e4}|^2\sin^2 (\Delta m^2_{41} L/E)+ \nonumber \\ 
& & |U_{e5}|^2\sin^2 (\Delta
m^2_{51} L/E))+ \nonumber \\ 
& &  |U_{e4}|^2|U_{e5}|^2\sin^2 (\Delta m^2_{54} L/E)].
\end{eqnarray}
This assumes that contributions to disappearance from the $\mu$ and
$\tau$ elements of the mixing matrix ($U_{\mu 4}$, $U_{\mu 5}$, $U_{\tau 4}$, and $U_{\tau 5}$) are negligible. We note that the current global fit improves significantly in the case of two sterile neutrinos~\cite{kopp}.

Figure~\ref{wiggles} illustrates the $L/E$-dependent signal for example 3+1 and 3+2 oscillation signals after five years of running. The observation of an oscillation wave, featuring multiple peaks and valleys for currently favored values of $\Delta m^2$, 
makes this a highly compelling analysis. The wave also allows differentiation between 3+1 and 3+2 models in most oscillation scenarios. The 3+2 model-based oscillation probability
shown in Fig.~\ref{wiggles} utilizes the oscillation parameters given in Ref.~\cite{kopp}. These parameters represent the best fit of the world's appearance and disappearance data.

\begin{figure}[h!]
\begin{center}
\begin{tabular}{c}
\hspace{-.8cm}
\includegraphics[scale=.27]{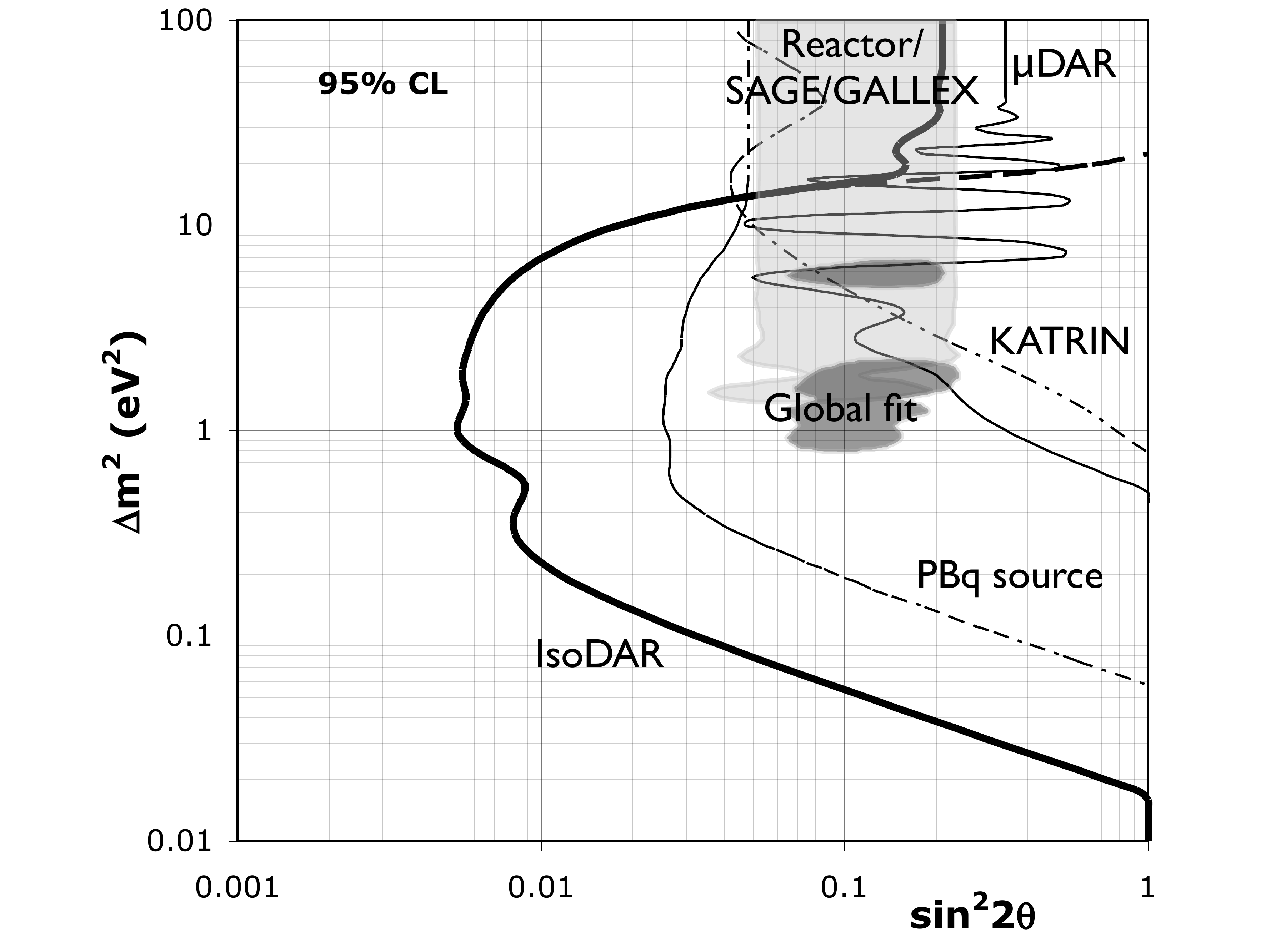}
\end{tabular}
\end{center}
\vspace{-.5cm}
\caption{The sensitivity of the IsoDAR experiment to electron antineutrino disappearance in a five~year physics run. The sensitivities for both rate+shape (solid line) and shape-only (dashed line) are shown. The $\mu$DAR~\cite{ConradShaevitz} exclusion curve and Reactor+Gallium~\cite{Mention} allowed region are also shown, along with the expected sensitivities from the PBq source~\cite{Thierry} and KATRIN~\cite{Formaggio} experiments.}
\label{fig:sensitivity}

\end{figure}

IsoDAR can quickly probe the oscillation parameter space indicative of one or more sterile neutrinos. As the antineutrino source described can be constructed within five years, we compare the IsoDAR 95\% CL sensitivity to experiments that can be accomplished on this time scale. The global fit region, encompassing all appearance and disappearance measurements, is shown along with this comparison in Fig.~\ref{fig:sensitivity}. Note that the global fit~\cite{kopp} pulls the reactor anomaly allowed region significantly lower in $\Delta m^2$ due to the LSND and MiniBooNE appearance results, resulting in a $\Delta m^2$$\sim$1-2~eV$^2$. As can be seen in Fig.~\ref{fig:sensitivity}, the statistics-limited IsoDAR sensitivity covers the 3+1 allowed
range [$\sin^2 2\theta=0.067$ and $\Delta m^2=1~\mathrm{eV}^2$] at 20$\sigma$ in five~years of running. IsoDAR can rule out $\sin^2 2\theta$=0.067, $\Delta m^2=1~\mathrm{eV}^2$ at 5$\sigma$ in 4~months. The ``shape-only" and ``rate+shape" based sensitivities are shown in the plot. It is clear that the flux normalization uncertainty is only relevant for oscillation sensitivity at high $\Delta m^2$ ($\gtrsim$$15~\mathrm{eV}^2$), a region where the rapid oscillation wave becomes difficult to resolve.  

The IsoDAR technique provides a high-intensity, low-energy source of antineutrinos with sensitivity to antineutrino oscillations near $\Delta m^2 \sim 1~\mathrm{eV}^2$. The experiment can perform compelling tests of models for new
physics that explain high $\Delta m^2$ oscillations through the
introduction of one or more sterile neutrinos. In a 3+1 model, IsoDAR
can address the global fit region for electron flavor disappearance at
20$\sigma$ (5$\sigma$) in five~years (four~months). In addition, the form of the oscillation wave can be reconstructed,
allowing differentiation between the existence of 3+1 and 3+2 neutrinos. The large event sample also provides the possibility of a wide variety of other standard model tests, including a precise measurement of the weak mixing angle. 

\clearpage
\begin{center}
{ \textbf{Acknowledgments}}
\end{center}
The authors thank the attendees of the Erice International School
of Subnuclear Physics Workshop, Nov. 27-Dec. 5, 2011, for valuable discussions.
Support for this workshop was provided through the Majorana Centre from the INFN Eloisatron Project directed by Prof. Antonino Zichichi. JMC and MHS thank the National Science Foundation for support.

\end{document}